\documentclass{ifacconf}

\usepackage{natbib}        
\usepackage{amsmath, amssymb}
\usepackage{color,soul}
\usepackage{graphicx}      
\usepackage{lipsum}
\usepackage{xcolor}
\usepackage{bm}
\usepackage{url}

\newcommand{\sss}[1]{_{\scriptscriptstyle #1}}

\newcommand\scalemath[2]{\scalebox{#1}{\mbox{\ensuremath{\displaystyle #2}}}}

\makeatletter
\def\endfigure{\end@float}
\def\endtable{\end@float}
\makeatother
\let\ifacconfcaptionwidth\captionwidth
\usepackage[caption=false]{subfig}
\let\captionwidth\ifacconfcaptionwidth

\begin{document}
\begin{frontmatter}

\title{Stability of discrete-time feed-forward neural networks in NARX configuration$^{*}$} 

\author{Fabio Bonassi}, 
\author{Marcello Farina}, 
\author{Riccardo Scattolini}
\thanks[*]{© 2021 the authors. This work has been accepted to IFAC for publication under a Creative Commons Licence CC-BY-NC-ND. Published version \citep{BONASSI2021547} available at \url{https://doi.org/10.1016/j.ifacol.2021.08.417}}

\address{Politecnico di Milano, Dipartimento di Elettronica, Informazione e
	Bioingegneria, Via Ponzio 34/5, 20133 Milano, Italy \\
	(e-mail: name.surname@polimi.it)}

\begin{abstract}
	The idea of using Feed-Forward Neural Networks (FFNNs) as regression functions for Nonlinear AutoRegressive eXogenous (NARX) models, leading to models herein named Neural NARXs (NNARXs), has been quite popular in the early days of machine learning applied to nonlinear system identification, owing to their simple structure and ease of application to control design.
	Nonetheless, few theoretical results are available concerning the stability properties of these models.
	In this paper we address this problem, providing a sufficient condition under which NNARX models are guaranteed to enjoy the Input-to-State Stability (ISS) and the Incremental Input-to-State Stability ($\delta$ISS) properties.
	This condition, which is an inequality on the weights of the underlying FFNN, can be enforced during the training procedure to ensure the stability of the model.
	The proposed model, along with this stability condition, are tested on the \emph{pH} neutralization process benchmark, showing satisfactory results. 
\end{abstract}


\end{frontmatter}

\section{Introduction}
The identification of nonlinear systems is a notoriously hard task.
Unlike the linear case, as thoroughly discussed by \cite{schoukens2019nonlinear}, nonlinear identification requires an accurate design of experiment, a suitable class of parametric nonlinear models, and an optimization-based numerical procedure to spot the best parametrization to fit the model.

Among the most popular model structures, Neural Networks (NNs) have lately risen as powerful tools to identify nonlinear systems \citep{forgione2020model}.
Traditional approaches, such as \cite{levin1995identification}, rely on Neural Nonlinear Auto Regressive eXogenous (NNARX), i.e. NARX models where Feed-Forward Neural Networks (FFNNs) are used as regression functions to predict the future output vector based on the past input and output data.
While these models typically enjoy satisfactory modeling performances, the more sophisticated Recurrent Neural Networks (RNNs) have been recently introduced for time-series forecasting and nonlinear system identification.
In particular, the most popular and promising recurrent architectures are Echo State Networks (ESNs, \cite{jaeger2002tutorial}), Long-Short Term Memory networks (LSTMs, \cite{hochreiter1997long}) and Gated Recurrent Units (GRUs, \cite{cho2014learning}), see \cite{bianchi2017overview} for a comparison. 

When these networks are used for model-based control strategies, however, the superior modeling performances of ESNs, LSTMs, GRUs, and RNNs in general, come at the price of more involved control algorithms. 
This is due not only to their complex structure, but also to the need of state-observers to operate these RNNs models in a closed-loop fashion, using the past input and output data to estimate the current states and improve the future outputs' predictions, as discussed by \cite{terzi2021learning}.

In the following we focus our attention on NNARX models, which -- unlike RNNs -- do not require state observers, as the future output depends on the known past inputs and outputs only \citep{levin1995identification}. 
For this reason, and in light of their simple structure and training, these models have been extensively applied for system identification and control, both in academia \citep{levin1993control, levin1996control} and in industry, in particular in chemical process control \citep{ali2015artificial, himmelblau2008accounts}.
Indeed, owing to their versatility, NNARXs have been widely employed with Model Predictive Control, showing remarkable results \citep{hosen2011control, nagy2007model, atuonwu2010identification}.

Despite the popularity of these models, limited research efforts have been devoted to the theoretical analysis of NNARXs. \cite{sanchez1999input} studied the Input-to-State Stability (ISS, \cite{jiang2001input}) of continuous-time NNARXs, retrieving a sufficient condition that network's weights must satisfy to guarantee such property. 
On the contrary, results on ISS and Incremental Input-to-State Stability ($\delta$ISS, \cite{bayer2013discrete}) properties have been recently published for RNNs \citep{miller2018stable}, ESNs \citep{armenio2019model}, LSTMs \citep{terzi2021learning, bonassi2019lstm}, GRUs \citep{bonassi2020stability, stipanovic2020stability}.
Both ISS and $\delta$ISS are required, among other applications, for the safety verification of the network \citep{bonassi2019lstm}, robust MPC design \citep{bayer2013discrete}, offset-free tracking MPC \citep{bonassi2021nonlinear}, and for Moving Horizon Estimators design \citep{alessandri2008moving}.

The goal of this paper is to fill this theoretical gap and devise conditions under which the NNARX models are guaranteed to enjoy ISS and $\delta$ISS.
In particular we devise a sufficient condition, in the form of a nonlinear inequality on network's weights, extending \citep{sanchez1999input} to discrete-time NNARXs, additionally providing results on the $\delta$ISS property.
This inequality can be either used to check a-posteriori the ISS and $\delta$ISS of a trained NNARX, or can be implemented in the training procedure to ensure the stability of the model that is being trained.
The resulting training procedure is discussed on a \emph{pH} neutralization process benchmark system \citep{hall1989modelling}, showing good modeling performances.

\subsection{Notation}
Given a vector $v$, we denote by $v^\prime$ its transpose and by $\| v \|$ its Euclidean norm.
Boldface fonts denote sequences of vectors, i.e. $\bm{v} = \{ v(0), v(1), ... \}$, where $\| \bm{v} \|_\infty = \max_{k \geq 0} \| v(k) \|$.
For compactness, when referring to a time-varying quantity, the time index $k$ is indicated in the subscript, e.g. $v_k$.

\section{NNARX models}
In NARX models the output $y\sss{k+1}$ is computed as a nonlinear regression over the past $N$ input and output samples, as well as the current input $u\sss{k}$ \citep{schoukens2019nonlinear}.
This definition can be represented as
\begin{equation} \label{eq:model:narx}
	y\sss{k+1} = f(y\sss{k}, y\sss{k-1}, ..., y\sss{k-N+1}, u\sss{k}, u\sss{k-1}, ..., u\sss{k-N}),
\end{equation}
where $f$ is the non-linear regression function. 
This function is assumed to be a parametrized vector function of $p$ Lipschitz-continuous functions, $p$ being the number system's outputs.

While $f$ can be any arbitrary regression function, such as polynomial functions \citep{piroddi2003identification}, in this work we consider Neural NARX models, in which a FFNN constitutes the nonlinear regression function.
Therefore, let us re-formulate the generic model \eqref{eq:model:narx} as a discrete-time system in a normal canonical form \citep{califano1998discrete}.
To this end, we define the following state vector containing the past data
\begin{equation} \label{eq:model:states_definition}
	z\sss{i, k} = \left[\begin{array}{c}
		y\sss{k-N+i} \\
		u\sss{k-N-1+i}
	\end{array}\right],
\end{equation}
with $i \in \{1, ..., N \}$. It is worth noticing that
\begin{equation*}
	z\sss{N, k+1} = \left[\begin{array}{c}
		y\sss{k+1} \\
		u\sss{k}
	\end{array}\right].
\end{equation*}
In light of \eqref{eq:model:states_definition}, it is possible to rewrite \eqref{eq:model:narx} as 
\begin{equation} \label{eq:model:normal_form}
	\left\{ \begin{array}{l}
		z\sss{1, k+1} = z\sss{2, k} \\
		z\sss{2, k+1} = z\sss{3, k} \\
		\quad \vdots \\
		z\sss{N-1,k+1} = z\sss{N, k} \\
		z\sss{N, k+1} = \begin{bmatrix}
			f(z\sss{1, k}, z\sss{2, k}, ..., z\sss{N, k}, u\sss{k}) \\
			u\sss{k}
		\end{bmatrix} \\
	y\sss{k} = [I\quad 0] \, z\sss{N, k}
	\end{array} \right.
\end{equation}
which is a discrete-time normal canonical form. Defining the state as $x\sss{k} = [ z\sss{1, k}^\prime, ..., z\sss{N, k}^\prime]^\prime$, \eqref{eq:model:normal_form} reads as
\begin{subequations} \label{eq:model:nnarx}
\begin{equation}
\begin{aligned} \label{eq:model:statespace}
	x\sss{k+1} &= \underbrace{\begin{bmatrix}
		0 & I & 0 & ... & 0 \\
		0 & 0 & I & ... & 0 \\
		\vdots &&& \ddots & \vdots \\
		0 & 0 & 0 & ... & I \\
		0 & 0 & 0 & ... & 0
	\end{bmatrix}}_{A} x\sss{k} +  
   \underbrace{\begin{bmatrix}
		0 \\
		0 \\
		\vdots \\
		0 \\
		\tilde{B}_u
	\end{bmatrix}}_{B_u} u\sss{k} +
	\underbrace{\begin{bmatrix}
		0 \\
		0 \\
		\vdots \\
		0 \\
		\tilde{B}_x
	\end{bmatrix}}_{B_x} f(x\sss{k}, u\sss{k}), \\
	y\sss{k} &= \underbrace{\begin{bmatrix}
			0 & ... & 0 & \tilde{C}
	\end{bmatrix}}_{C} x\sss{k}
\end{aligned}
\end{equation}
where $0$ and $I$ are null and identity matrices of proper dimensions. In particular, denoting by $m$ the number of inputs, each block appearing in $A$ has dimension $(m+p) \times (m+p)$, while the blocks appearing in $B_u$ and $B_x$ are $(m+p) \times m$ and $(m+p) \times p$ matrices, respectively. 
The overall dimension of $A$ is hence $n \times n$, where $n=(m+p) N$.
The blocks appearing in matrix $C$ are $p \times (m+p)$.
The sub-matrices $\tilde{B}_u$, $\tilde{B}_{x}$ and $\tilde{C}$ are defined as
\begin{equation*}
	\tilde{B}_u = \begin{bmatrix}
		0_{p\times m} \\
		I_{m \times m}
	\end{bmatrix}, \quad
	\tilde{B}_x = \begin{bmatrix}
		I_{p \times p} \\
		0_{m \times p}
	\end{bmatrix}, \quad
	\tilde{C} = \begin{bmatrix}
		I_{p \times p} & 0_{p \times m}
	\end{bmatrix}.
\end{equation*}

For Neural NARX models, i.e. NNARXs, the function $f$ is realized by means of a feed-forward neural network.
Such networks are static maps consisting of $M$ layers of neurons, each layer being a linear combination of its inputs, passed through an appropriate nonlinear function named activation function. 
The network can be compactly written as
	\begin{equation}  \label{eq:model:ffnn}
		f(x\sss{k}, u\sss{k}) = U_0 \, f_M(f_{M-1}(... f_1(x\sss{k}, u\sss{k}), u\sss{k}), u\sss{k}) + b_0,
	\end{equation}
where $f_i$ is the nonlinear relation established by the $i$-th layer, which reads as
\begin{equation}\label{eq:model:ffnn_layer}
	f_i(f_{i-1}, u\sss{k}) = \sigma_i \big( W_i u\sss{k} + U_i f_{i-1} + b_i \big),
\end{equation}
\end{subequations}
$f_{i-1}$ being the output of the previous layer, or $x\sss{k}$ if $i=1$.
Each layer is parametrized by the matrices $W_i$, $U_i$, and $b_i$, and its activation function $\sigma_i$ is assumed to be zero-centered ($\sigma_i(0) = 0$) and Lipschitz continuous by a Lipschitz constant $L_{\sigma_i}$. For example, one may take $\sigma_i = \text{tanh}$, in which case $L_{\sigma_i}=1$.

To summarize, NNARX models in state-space form \eqref{eq:model:statespace} are considered, where the nonlinear regression function is described by the feed-forward NN \eqref{eq:model:ffnn}-\eqref{eq:model:ffnn_layer}.



\section{Stability properties}
The goal of this section is to provide conditions under which the NNARX models \eqref{eq:model:nnarx} are guaranteed to enjoy the ISS and $\delta$ISS properties.

For compactness, in the following we denote by $x\sss{k}(\bar{x}, \bm{u}, b)$ the state at time $k$, starting from the initial condition $\bar{x}$, when the system is fed by the input sequence $\bm{u}$, and where the vector of biases is $b = [ b_0^\prime, b_1^\prime, ..., b_M^\prime]^\prime$. 
The reason for which the biases, differently from other system's parameters $W_i$ and $U_i$, are explicitly indicated here is that they act additively in the network's activation functions.
Therefore, as better explained in the following, they can be regarded, roughly speaking, as constant inputs.

\subsection{Input-to-State Stability}
Recalling the definition of $\mathcal{KL}$ and $\mathcal{K}_\infty$ functions from \cite{jiang2001input}, the following definition are given.

\begin{defn}[ISS] \label{def:ISS}
	System \eqref{eq:model:nnarx} is Input-to-State Stable (ISS) if there exist functions $\beta( \| \bar{x} \|, k) \in \mathcal{KL}$, $\gamma_u(\| \bm{u} \|_\infty ) \in \mathcal{K}_\infty$, and $\gamma_b(\| b_r \| ) \in \mathcal{K}_\infty$, such that for any $k \in \mathbb{Z}_{\geq 0}$, any initial condition $\bar{x}$, any value of $b$, and any input sequence $\bm{u}$, it holds that 
	\begin{equation} \label{eq:def:ISS}
		\| x\sss{k}(\bar{x}, \bm{u}, b) \| \leq \beta(\| \bar{x} \|, k) +\gamma_u(\| \bm{u} \|_\infty )  + \gamma_b(\| b \| ).
	\end{equation}
\end{defn}

\vspace{1mm}

\begin{defn}[ISS-Lyapunov function]
	A continuous function $V: \mathbb{R}^n \to \mathbb{R_+}$ is said to be an ISS-Lyapunov function for system \eqref{eq:model:nnarx} if there exist functions $\psi\sss{1}, \psi\sss{2}, \psi, \varphi\sss{u}, \varphi\sss{b} \in \mathcal{K}_\infty$ such that, for all $x\sss{k}$, $u\sss{k}$, and $b$, it holds that
	\begin{equation} \label{eq:iss:lyapunov}
		\begin{gathered}
			\psi\sss{1} (\| x\sss{k} \|) \leq V( x\sss{k} ) \leq \psi\sss{2} (\| x\sss{k} \|), \\
			V( x\sss{k+1} ) - V ( x\sss{k} ) \leq -\psi(\| x\sss{k} \|) + \varphi\sss{u}(\| u\sss{k} \|) + \varphi\sss{b}(\| b \|),
		\end{gathered}
	\end{equation}
	where $x\sss{k+1}$ is determined by \eqref{eq:model:nnarx}.
\end{defn}

Then, in light of the following Lemma, finding an ISS-Lyapunov function for the system allows to assess the ISS property of the system

\begin{lem}[Lemma 3.5, \cite{jiang2001input}] \label{lemma:iss}
	If system \eqref{eq:model:nnarx} admits a continuous ISS-Lyapunov function, it is ISS. 
\end{lem}

Under these premises, the following theoretical contribution can be stated.

\begin{thm} \label{thm:iss}
	A sufficient condition for the ISS of the NNARX model \eqref{eq:model:nnarx} is that
	\begin{equation} \label{eq:iss:condition}
		\prod_{i=0}^{M} \| U_i \| < \frac{1}{\big(\prod_{i=1}^{M} L_{\sigma i}\big) \sqrt{N}}.
	\end{equation}
\end{thm}

\begin{pf}
	Define $P = \text{diag}(I, 2 \cdot I, ..., N \cdot I)$. It is easy to see that $P$ is the solution to the Lyapunov equation $A^\prime P A - P = -Q$, when $Q=I$. 
	Let us then consider the candidate ISS-Lyapunov function $V(x) = x^{\prime} P x$.
	It holds that
	\begin{equation}
		 \| x\sss{k} \|^2 \leq V(x\sss{k}) \leq N \| x\sss{k} \|^2,
	\end{equation}
	hence in \eqref{eq:iss:lyapunov} $\psi\sss{1}(\| x\sss{k} \|) = \| x\sss{k} \|^2$ and $\psi\sss{2} (\| x\sss{k} \|) = N \| x\sss{k} \|^2$. Furthermore,
	\begin{equation} \label{eq:iss:proof:Delta_V}
		\begin{aligned}
			& V(x\sss{k+1}) - V(x\sss{k}) = x\sss{k+1}^\prime P x\sss{k+1} - x\sss{k}^\prime P x\sss{k} \\
			& \quad = x\sss{k}^\prime ( A^\prime P A - P) x\sss{k} + u\sss{k}^\prime B_u^\prime P B_u u\sss{k} + 2 x\sss{k}^\prime A^\prime P B_u u\sss{k} \\
			& \qquad + f(x\sss{k}, u\sss{k})^\prime B_x^\prime  P B_x f(x\sss{k}, u\sss{k}) + 2 x\sss{k}^\prime A^\prime P B_x f(x\sss{k}, u\sss{k}) \\
			& \qquad + 2  u\sss{k}^\prime B_u^\prime P B_x f(x\sss{k}, u\sss{k}).
		\end{aligned}
	\end{equation}
	In light of the structure of $A$, $B_u$ and $B_x$, and being $P$ block-diagonal, it follows that
	\begin{equation} \label{eq:iss:matrix_properties}
		\begin{aligned}
			& A^\prime P B_u = 0_{n \times m} \quad && A^\prime P B_x = 0_{n \times p}, \\
			& B_u^\prime P B_u = N \cdot I_{m \times m} \quad && B_x^\prime P B_x = N \cdot I_{p \times p}, \\
			& B_u^\prime P B_x = N \tilde{B}_u^\prime \tilde{B}_x = 0_{m \times p}. &&
		\end{aligned}
	\end{equation}
	Equation \eqref{eq:iss:proof:Delta_V} can hence be rewritten as
	\begin{equation} \label{eq:iss:proof:Delta_V_red}
		V(x\sss{k+1}) - V(x\sss{k}) = - x\sss{k}^\prime x\sss{k} + N u\sss{k}^\prime u\sss{k} + N f(x\sss{k}, u\sss{k})^\prime f(x\sss{k}, u\sss{k})
	\end{equation}
	Owing to the Lipschitzianity of $\sigma_i$, by standard norms arguments, for any $\alpha \neq 0$ it holds that
	\begin{equation} \label{eq:iss:proof:f_norm}
	\scalemath{1}{
		\begin{aligned}
			 \| f(x\sss{k}, u\sss{k}) \|^2 \leq& \Big(1+\frac{1}{\alpha^2}\Big) K_x^2 \| x\sss{k} \|^2 + 2 (1 +\alpha^2) K_u^2 \| u\sss{k} \|^2 \\
			&  +  2 (1 +\alpha^2) K_b^2 \| b_i \|^2
		\end{aligned}}
	\end{equation}
	where $K_x$, $K_u$, and $K_b$ are defined as
	\begin{equation} \label{eq:iss:proof:K_def}
	\scalemath{0.9}{
		\begin{aligned}
			 K_x =& \| U_0 \| \prod_{i=1}^{M} L_{\sigma i} \| U_i \|, \\
			 K_u =& \| U_0 \| \sum_{i=1}^{M}  \bigg(  \prod_{j=i+1}^{M} L_{\sigma_j} \| U_j \| \bigg)  L_{\sigma_i} \| W_i \|, \\
			 K_b =& \| U_0 \| \sum_{i=1}^{M}  \bigg(  \prod_{j=i+1}^{M} L_{\sigma_j} \| U_j \| \bigg)  L_{\sigma_i}.
		\end{aligned}}
	\end{equation}
	Combining \eqref{eq:iss:proof:Delta_V_red} and \eqref{eq:iss:proof:f_norm} we get
	\begin{equation}
		\scalemath{0.9}{
		\begin{aligned}
			&V(x\sss{k+1}) - V(x\sss{k}) \leq - \Big[ 1 - \Big( 1 + \frac{1}{\alpha^2} \Big) N K_x^2 \Big] \| x\sss{k} \|^2 \\
			&\quad + N \bigg[ 1 + 2(1+\alpha^2) K_u^2 \bigg] \| u\sss{k} \|^2  + 2(1+\alpha^2) N K_b^2 \| b \|^2
		\end{aligned}}
	\end{equation}

	$V$ is an ISS-Lyapunov function if the coefficient multiplying $\| x\sss{k} \|^2$ is strictly negative, which holds if
	\begin{equation} \label{eq:iss:proof:condition_alpha}
		\prod_{i=0}^{M} \| U_i \| < \frac{1}{\big(\prod_{i=1}^{M} L_{\sigma i}\big) \sqrt{N}} \sqrt{\frac{\alpha^2}{1+ \alpha^2}}.
	\end{equation}
	
	If \eqref{eq:iss:condition} holds, then for a sufficiently large value of $\alpha$ there exists $\varepsilon > 0$ such that
	 \begin{equation}
	 \scalemath{0.9}{
	 \begin{aligned}
	 	\prod_{i=0}^{M} \| U_i \| < \frac{1}{\big(\prod_{i=1}^{M} L_{\sigma i}\big) \sqrt{N}} (1-\varepsilon) < \frac{1}{\big(\prod_{i=1}^{M} L_{\sigma i}\big) \sqrt{N}} \sqrt{\frac{\alpha^2}{1+ \alpha^2}}.
	 \end{aligned}}
	\end{equation}
	 Therefore \eqref{eq:iss:proof:condition_alpha} is satisfied and, in light of \eqref{eq:iss:proof:K_def}, there exists $\delta > 0$ such that
	\begin{equation}
		- \bigg( 1 - \Big( 1 + \frac{1}{\alpha^2} \Big) N K_x^2 \bigg) < - \delta,
	\end{equation}
	hence $V$ is an ISS-Lyapunov function, with functions 
	$ \psi(\| x\sss{k} \|) \! =\! -\delta \| x\sss{k} \|^2$, $\varphi\sss{u}(\| u\sss{k} \|_\infty) \! = \! N [ 1 + 2(1+\alpha^2) K_u^2 ] \| u\sss{k} \|^2$ and $\varphi\sss{b}(\| b \|) =  2(1+\alpha^2) N K_b^2 \| b \|^2$.
	In light of Lemma \ref{lemma:iss}, system \eqref{eq:model:nnarx} is ISS. $\hfill \blacksquare$
\end{pf}

\subsection{Incremental Input-to-State Stability}
In this section the following notation is adopted for the sake of compactness. We indicate a pair of generic initial states by $\bar{x}\sss{a}$ and $\bar{x}\sss{b}$, and a pair of generic input sequences by $\bm{u}\sss{a} = \{u\sss{a,0}, u\sss{a, 1}, ...\}$ and $\bm{u}\sss{b} = \{u\sss{b,0}, u\sss{b, 1}, ...\}$.
We denote by $x\sss{a,k} = x\sss{a,k}(\bar{x}\sss{a}, \bm{u}\sss{a}, b)$ the state trajectory at time $k$, obtained initializing system \eqref{eq:model:nnarx} in the initial state $\bar{x}\sss{a}$ and feeding it with the input sequence $\bm{u}\sss{a}$. The same notation is used for $x\sss{b,k} = x\sss{b,k}(\bar{x}\sss{b}, \bm{u}\sss{b}, b)$. 
The following definitions from \cite{bayer2013discrete} can hence be given. \vspace{1mm}

\begin{defn}[$\delta$ISS] \label{def:deltaISS}
	System \eqref{eq:model:nnarx} is Incrementally Input-to-State Stable ($\delta$ISS) if there exist functions $\beta( \| \bar{x}_a - \bar{x}_b \|, k) \in \mathcal{KL}$ and $\gamma_u(\| \bm{u}\sss{a} - \bm{u}\sss{b} \|_\infty ) \in \mathcal{K}_\infty$ such that for any $k \in \mathbb{Z}_{\geq 0}$, any pair of initial conditions $\bar{x}\sss{a}$ and $\bar{x}_b$, and any pair of input sequences $\bm{u}\sss{a}$ and $\bm{u}\sss{b}$, it holds that 
	\begin{equation} \label{eq:def:deltaISS}
		\scalemath{1}{
			\| x\sss{a, k} -  x\sss{b, k} \| \leq \beta(\| \bar{x}\sss{a} - \bar{x}\sss{b} \|, k) + \gamma_u(\| \bm{u}\sss{a} - \bm{u}\sss{b} \|_\infty ).}
	\end{equation}
\end{defn}

Note that the $\delta$ISS property implies that, initializing the network in different states and feeding it with different input sequences, one obtains state trajectories which are asymptotically bounded by a function $\gamma_u$ which is monotonically increasing with the maximum difference between the two input sequences.


\begin{defn}[$\delta$ISS-Lyapunov function]
	$\,\,$ A continuous function $V\sss{\delta}: \mathbb{R}^{n \times n} \to \mathbb{R_+}$ is said to be a $\delta$ISS-Lyapunov function for system \eqref{eq:model:nnarx} if there exist functions $\psi\sss{ 1}, \psi\sss{ 2}, \psi, \varphi \in \mathcal{K}_\infty$ such that, for any  $x\sss{a,k}$ and $x\sss{b,k}$, and any $u\sss{a,k}$ and $u\sss{b,k}$, it holds that \vspace{2mm}
	\begin{equation}
		\scalemath{0.8}{
		\begin{gathered}
			\psi\sss{1} (\| x\sss{a, k} - x\sss{b, k} \|) \leq V\sss{\delta} ( x\sss{a, k}, x\sss{b, k} ) \leq \psi\sss{2} (\| x\sss{a k} - x\sss{b k} \|), \\
			V\sss{\delta}( x\sss{a, k+1}, x\sss{b, k+1}  ) - V\sss{\delta} ( x\sss{a, k}, x\sss{b, k} ) \leq -\psi(\| x\sss{a, k} - x\sss{b, k} \|) + \varphi(\| u\sss{a, k} - u\sss{b, k} \|),
		\end{gathered}}
	\end{equation}
	where $x\sss{a, k+1}$ and $x\sss{b, k+1}$ are determined by \eqref{eq:model:nnarx}.
\end{defn}

Similarly to the ISS property, the existence of a $\delta$ISS-Lyapunov function is tied to the $\delta$ISS of the system through the following Lemma.

\begin{lem}[Theorem 1, \cite{bayer2013discrete}] \label{lemma:deltaISS}
	If it admits a continuous $\delta$ISS-Lyapunov function, system \eqref{eq:model:nnarx} is $\delta$ISS.
\end{lem}

In light of these definitions and of Lemma \ref{lemma:deltaISS}, the following Theorem can be stated.

\begin{thm} \label{thm:deltaiss}
	If system \eqref{eq:model:nnarx} is ISS by Theorem \ref{thm:iss}, i.e. condition \eqref{eq:iss:condition} is fulfilled, then it is also $\delta$ISS.
\end{thm}
\begin{pf}
	Consider $V\sss{\delta}(x\sss{a}, x\sss{b}) = (x\sss{a} - x\sss{b})^\prime P (x\sss{a} - x\sss{b})$ as a candidate $\delta$ISS-Lyapunov function, where $P$ is the solution to the Lyapunov equation $A^\prime P A - P = -Q$, with $Q=-I$.
	Then $P$ is a block-diagonal matrix, $P = P^\prime = \text{diag}(I, 2 \cdot I, ..., M \cdot I)$, where I is the $(m+p) \times (m+p)$ identity matrix.
	It holds that
	\begin{equation}
		\| x\sss{a, k} - x\sss{b, k} \|^2 \leq V\sss{\delta}(x\sss{a,k}, x\sss{b,k}) \leq N \| x\sss{a, k} - x\sss{b, k} \|^2,
	\end{equation}
	thus $\psi\sss{1}(\| x\sss{a,k} - x\sss{b,k} \|) =  \| x\sss{a,k} - x\sss{b,k} \|^2$ and $\psi\sss{2}(\| x\sss{a,k} - x\sss{b,k} \|) = N \| x\sss{a,k} - x\sss{b,k} \|^2$.

	Denoting by $\Delta V\sss{\delta,k} = V\sss{\delta}(x\sss{a,k+1}, x\sss{b,k+1}) - V\sss{\delta}(x\sss{a,k}, x\sss{b,k})$, it follows that
	\begin{equation}
	\scalemath{0.72}{
		\begin{aligned}
			& \Delta V\sss{\delta,k} = 	\big[ A x\sss{a, k} + B_u u\sss{a, k} + B_x f(x\sss{a, k}, u\sss{a, k}) - A x\sss{b, k} - B_u u\sss{b, k} - B_x f(x\sss{b, k}, u\sss{b, k}) \big]^\prime \cdot \\
			& \quad \cdot P \cdot \big[ A x\sss{a, k} + B_u u\sss{a, k} + B_x f(x\sss{a, k}, u\sss{a, k}) - A x\sss{b, k} - B_u u\sss{b, k} - B_x f(x\sss{b, k}, u\sss{b, k}) \big]  \\
			& \quad - (x\sss{a, k} - x\sss{b, k})^\prime P (x\sss{a, k} - x\sss{b, k}) 
		\end{aligned}}
	\end{equation}
	In light of \eqref{eq:iss:matrix_properties}, the previous equality can be re-written as
	\begin{equation}
		\scalemath{0.82}{
			\begin{aligned}
			& \Delta V\sss{\delta,k} = 	(x\sss{a, k} - x\sss{b, k})^\prime (A^\prime P A - P) (x\sss{a, k} - x\sss{b, k}) \\
			& \quad + (u\sss{a, k} - u\sss{b, k})^\prime B_u^\prime P B_u (u\sss{a, k} - u\sss{b, k}) \\
			& \quad + \big[ f(x\sss{a, k}, u\sss{a, k}) - f(x\sss{b, k}, u\sss{b, k}) \big]^\prime B_x^\prime P B_x \big[ f(x\sss{a, k}, u\sss{a, k}) - f(x\sss{b, k}, u\sss{b, k}) \big]
		\end{aligned}}
	\end{equation}
	
	By summing and subtracting $f(x\sss{b, k}, u\sss{a, k})$ to  the square brackets of the third term, and applying standard norm arguments, for any $\alpha \neq 0$ it holds that 
	\begin{equation} \label{eq:deltaiss:f_norms1}
		\scalemath{0.82}{
			\begin{aligned}
			& \big[ f(x\sss{a, k}, u\sss{a, k}) - f(x\sss{b, k}, u\sss{b, k}) \pm f(x\sss{b, k}, u\sss{a, k}) \big]^\prime B_x^\prime P \cdot \\
			& \quad  \cdot B_x \big[ f(x\sss{a, k}, u\sss{a, k}) - f(x\sss{b, k}, u\sss{b, k})  \pm f(x\sss{b, k}, u\sss{a, k}) \big] \\
			& \leq N \left\| \Big( f(x\sss{a, k}, u\sss{a, k}) - f(x\sss{b, k}, u\sss{a, k})\Big) + \Big( f(x\sss{b, k}, u\sss{a, k}) -  f(x\sss{b, k}, u\sss{b, k})\Big)\right\|^2 \\
			& \leq N \Big(1+\frac{1}{\alpha^2}\Big) \| f(x\sss{a, k}, u\sss{a, k}) - f(x\sss{b, k}, u\sss{a, k}) \|^2 \\
			& \quad + N \Big(1+\alpha^2\Big) \| f(x\sss{b, k}, u\sss{a, k}) - f(x\sss{b, k}, u\sss{b, k}) \|^2
		\end{aligned}}
	\end{equation}
	Then, since $f(x\sss{k}, u\sss{k})$ is Lipschitz continuous
	\begin{equation} \label{eq:deltaiss:f_norms2}
		\scalemath{1}{
			\begin{aligned}
			\| f(x\sss{a, k}, u\sss{a, k}) - f(x\sss{b, k}, u\sss{a, k}) \|^2 &\leq K_{x}^2 \| x\sss{a, k} - x\sss{b, k}\|^2, \\
			\| f(x\sss{b, k}, u\sss{a, k}) - f(x\sss{b, k}, u\sss{b, k}) \|^2 &\leq K_{u}^2 \| u\sss{a, k} - u\sss{b, k}\|^2,
		\end{aligned}}
	\end{equation}
	where $K_{x}$ and $K_{u}$ are defined as in \eqref{eq:iss:proof:K_def}.	
		In light of \eqref{eq:deltaiss:f_norms1} and \eqref{eq:deltaiss:f_norms2}, since $A^\prime P A - P = -Q$, with $Q=I$, $\Delta V\sss{\delta,k}$ can be re-formulated as
	\begin{equation} \label{eq:deltaiss:deltav_final}
		\scalemath{1}{
			\begin{aligned}
			 \Delta V\sss{\delta,k} \leq& -\Big[ 1 -   \Big(1 + \frac{1}{\alpha^2} \Big) N K_x^2\Big] \| x\sss{a,k} - x\sss{b,k}\|^2 \\&+ N \Big[ 1 + (1 + \alpha^2) K_u^2 \Big] \| u\sss{a, k} - u\sss{b, k}\|^2 
		\end{aligned}}
	\end{equation}
	Therefore, $V\sss{\delta}$ is a $\delta$ISS-Lyapunov function provided that the coefficient multiplying $\| x\sss{a,k} - x\sss{b,k}\|^2$ is negative, i.e. if there exists $\alpha \neq 0$ such that
	\begin{equation} \label{eq:deltaiss:proof_cond}
		\prod_{i=0}^M \| U_i \| < \frac{1}{\big( \prod_{i=1}^M   L_{\sigma i} \big) \sqrt{N}} \, \sqrt{\frac{\alpha^2}{1+\alpha^2}}.
	\end{equation}
	As discussed in the proof of Theorem \ref{thm:iss}, if Assumption \eqref{eq:iss:condition} holds, for $\alpha$ sufficiently large there exist  $\delta > 0$ such that $-\big[   1 - \big(1 + \frac{1}{\alpha^2} \big) N K_x^2 \big] < -\delta$. Hence, $V\sss{\delta}$ is a $\delta$ISS-Lyapunov function, with $\psi(\| x\sss{a, k} - x\sss{b, k} \|) = - \delta \| x\sss{a, k} - x\sss{b, k} \|^2$ and $\varphi(\| u\sss{a, k} - u\sss{b, k} \|) = N[ 1 + (1+ \alpha^2) K_u^2 ] \| u\sss{a, k} - u\sss{b, k} \|^2$.
		By Lemma \ref{lemma:deltaISS} system \eqref{eq:model:nnarx} is $\delta$ISS. $\hfill \blacksquare$
\end{pf}

\subsection{Summary}
	Theorem \ref{thm:iss} and Theorem \ref{thm:deltaiss} provide sufficient conditions that the weights of the feed-forward neural network \eqref{eq:model:ffnn} must satisfy to ensure the ISS and $\delta$ISS of the NNARX model \eqref{eq:model:nnarx}.
	These conditions boil down to a single nonlinear inequality, which can be either used to a-posteriori certify the ISS and $\delta$ISS of a trained NNARX, or can be employed during the training procedure to ensure the stability of the model.
	In particular, if the training algorithm does not support explicit constraints, condition \eqref{eq:iss:condition} can be relaxed and suitably accounted in the cost function, {as discussed by} \cite{bonassi2020stability}.

\section{Numerical results}
\begin{figure}
	\centering
	\includegraphics[width=0.55 \linewidth]{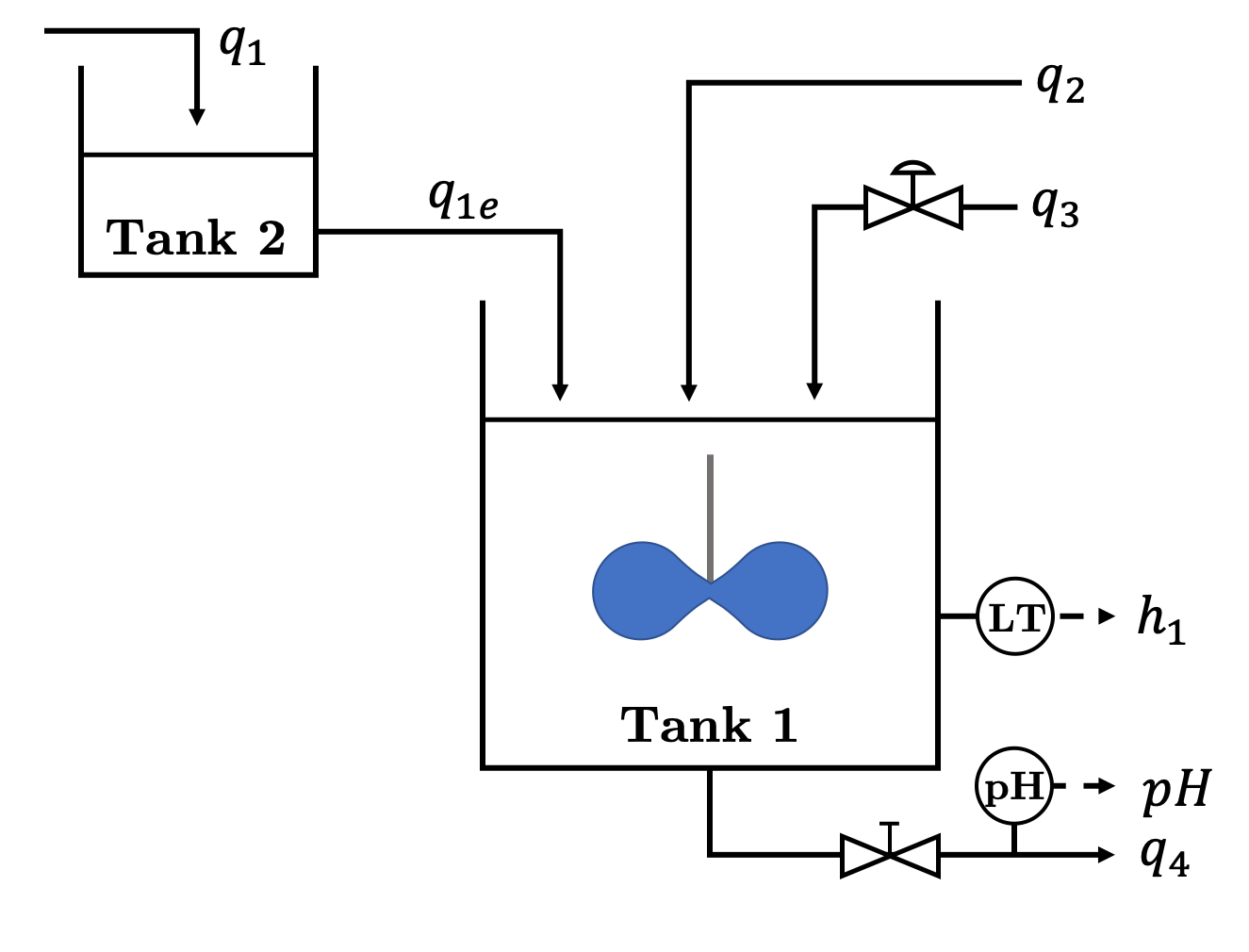}
	\vspace{-3mm}
	\caption{Scheme of the \emph{pH} neutralization process system.}
	\label{fig:ph}
\end{figure}

The proposed approach has been tested on the \emph{pH} neutralization process described by \cite{hall1989modelling} and considered in \cite{bonassi2019lstm} and \cite{terzi2021learning}. 
The plant, schematically represented in Figure \ref{fig:ph}, features two tanks. 
Tank $2$ is characterized by the acid flow rate $q_1$ as input and the flow rate $q_{1e}$ as output. 
It is assumed that the hydraulic dynamics are fast enough that $q_1 = q_{1e}$.
Tank $1$ is fed by three inputs, namely $q_1$, an uncontrollable buffer flow rate $q_2$, and an alkaline flow rate $q_3$ modulated by a controllable valve.
The \emph{pH} of the output flow rate of Tank $1$, i.e. $q_4$, is measured.
The overall simplified model is a third-order nonlinear SISO system, where the controllable input is the alkaline flow rate $q_3$ and the measured output is the \emph{pH}. 
This model is reported in \cite{terzi2021learning}.

\begin{figure}
	\centering
	\subfloat[]{\includegraphics[width=0.425\linewidth]{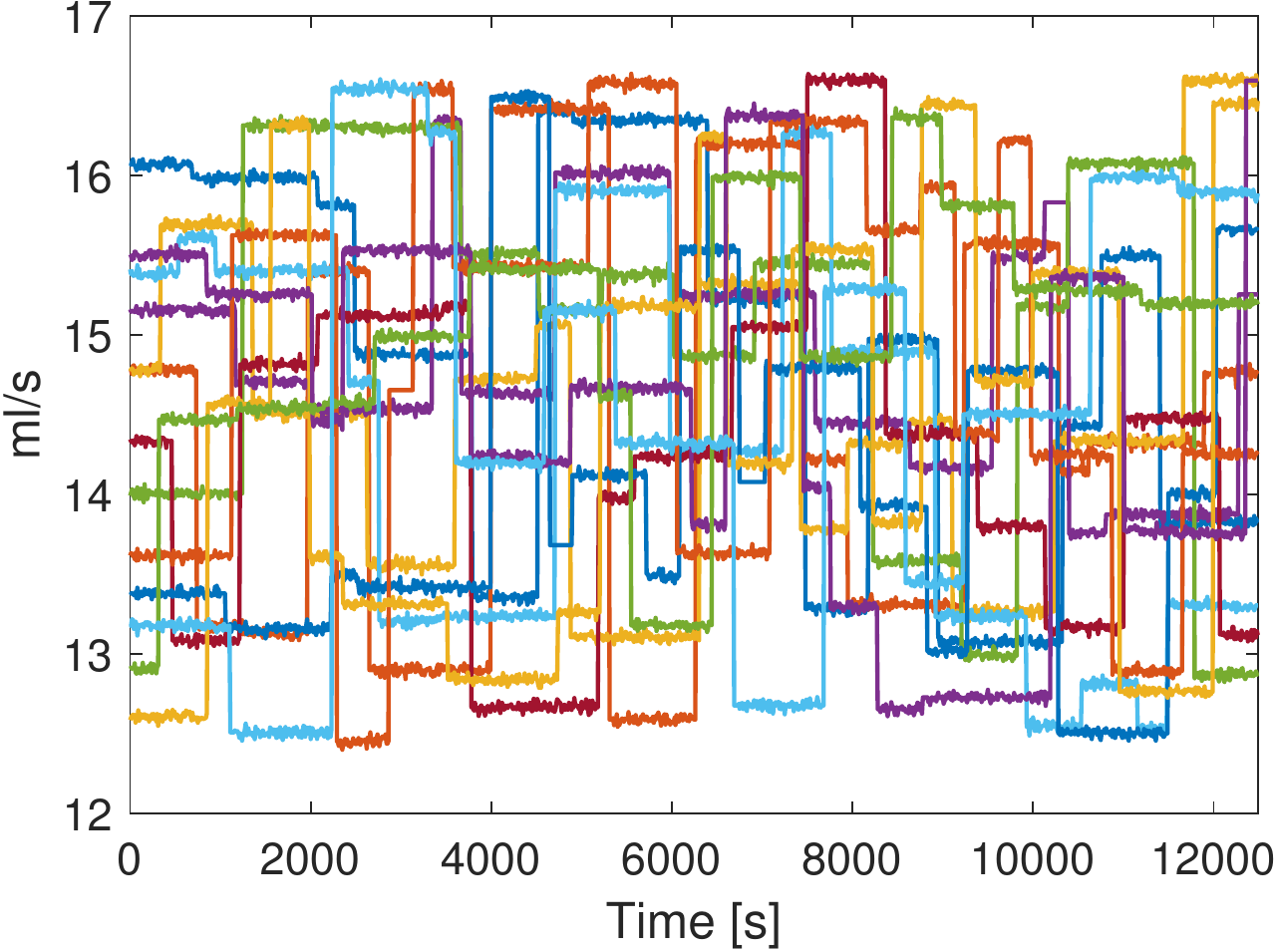}}\hspace{3mm}
	\subfloat[]{\includegraphics[width=0.425\linewidth]{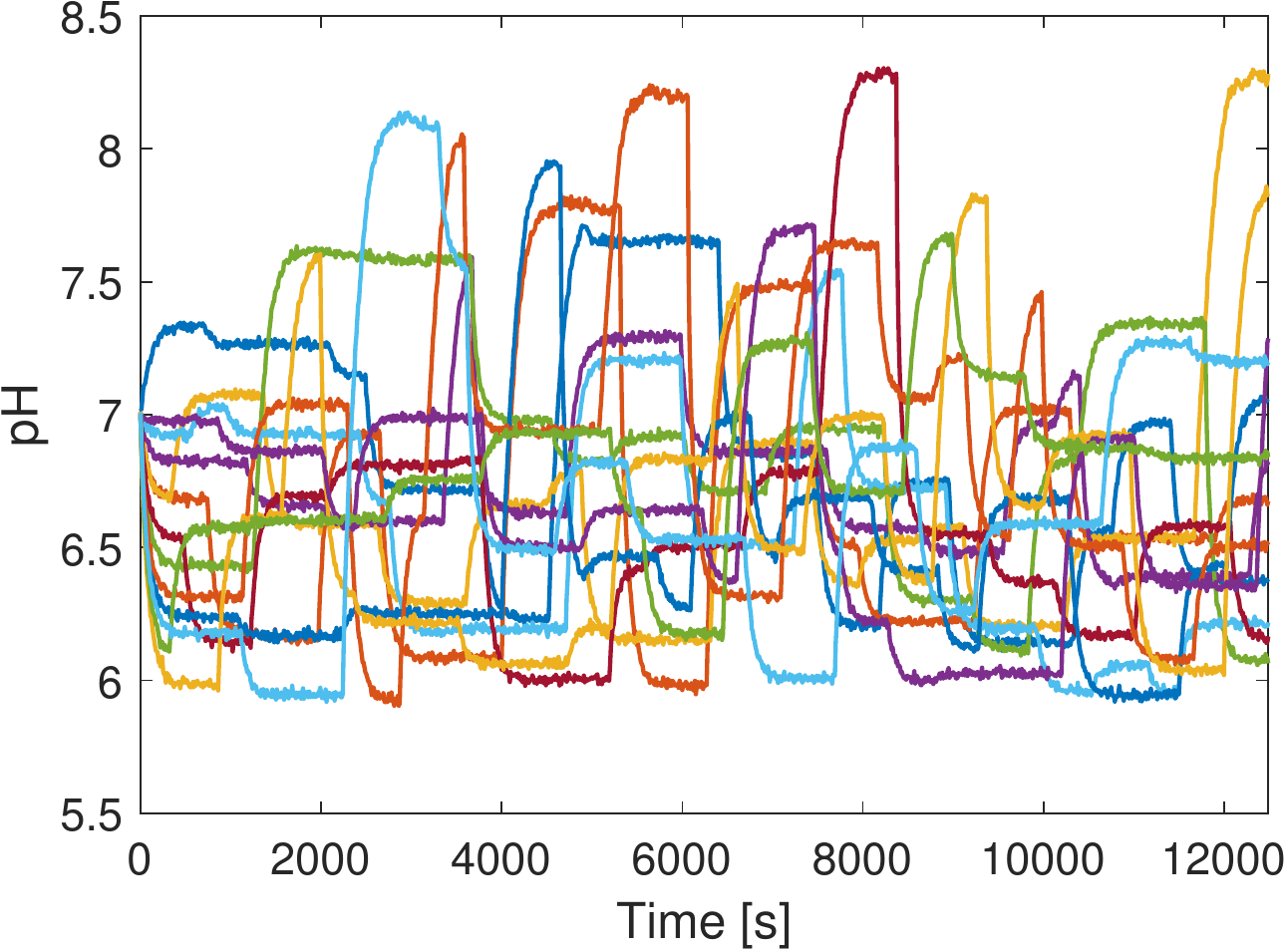}}
	\vspace{-3mm}
	\caption{Dataset used for training and validation of the NNARX model: (a) input and (b) output trajectories.}
	\label{fig:dataset}
\end{figure}

To generate the dataset used for the training of the NNARX model, a simulator of the system was implemented in MATLAB, and it was fed with Multilevel Pseudo-Random Signals (MPRS) in order to properly excite the system in a broad operating region.
A total of $N_s = 13$ input-output trajectories $(\bm{u}^{\{i\}}, \bm{y}^{\{i\}})$, with $i \in \{1, ..., N_s \}$, were collected with a sampling time $T_s = 10 s$. Each trajectory consists of $T_s = 1250$ samples $(u\sss{k}, y\sss{k})$. 
White noise was added both to the input and to the output to mitigate overfitting. The collected input and output trajectories, depicted in Figure \ref{fig:dataset}, were divided in $10$ trajectories for training and $3$ for validation.
The data was then normalized with respect to the mean and maximum deviation to ease the training.

The adopted NNARX model is characterized by a single-layer ($M=1$) FFNN with $10$ units and activation function $\sigma_1 = \text{tanh}$. 
The chosen look-back horizon is $N=4$.
Thus, being $m=p=1$ and $n = (m+p)N = 8$, the weight matrices have the following dimensions: $W_1$ is a $10 \times 1$; $U_1$ is a $10 \times 8$; $b_1$ is a $10 \times 1$; $U_0$ is a $1 \times 10$; $b_0$ is a scalar.
The training procedure was conducted using TensorFlow 1.15 on Python 3.7. As discussed in \cite{bonassi2020stability}, since TensorFlow does not support constrained training, the stability condition \eqref{eq:iss:condition} was enforced by means of a suitable regularization term $\rho(\nu)$ in the loss function:
\begin{equation} \label{eq:example:loss}
	L = \frac{1}{T_s - T_{w}} \sum_{k = T_{w}}^{T_s} \Big( y\sss{k}(\bar{x}, \bm{u}^{\{i\}}) - y\sss{k}^{\{i\}} \Big)^2 + \rho(\nu),
\end{equation}
where $y_k(\bar{x}, \bm{u}^{\{i\}})$ denotes the output of the NNARX model \eqref{eq:model:nnarx}, initialized in the random state $\bar{x}$ and fed by the input sequence $\bm{u}^{\{i\}}$. 
The initial $T_{w}$ data points are discarded to accommodate the effect of the initialization.
Note that during training, the NNARX model is used to perform an open-loop simulation throughout the entire trajectory, in an Output-Error fashion \citep{schoukens2019nonlinear}.
The regularization term is designed to penalize the residual of constraint \eqref{eq:iss:condition}, i.e.
\begin{equation} \label{eq:example:residual}
	\nu = \prod_{i=0}^M \| U_i \| - \frac{1}{\prod_{i=1}^M L_{\sigma i} \sqrt{N}}.
\end{equation}
Note that when $\nu < 0$ the stability condition is fulfilled. Hence, a simple piece-wise linear function can be adopted for $\rho(\nu)$  \citep{bonassi2020stability}.

\begin{figure}
	\centering
	\includegraphics[width=0.625 \linewidth]{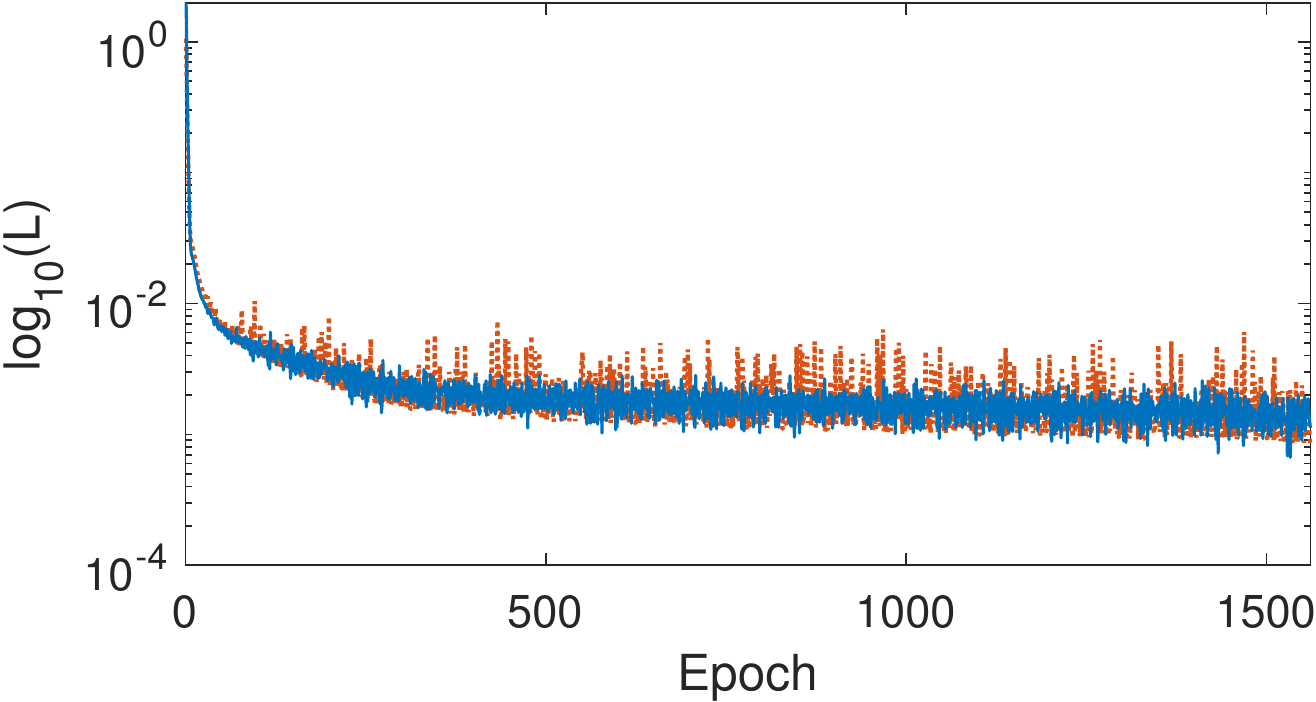}
	\vspace{-3mm}
	\caption{Evolution of the loss function $L$ throughout the training procedure.}
	\label{fig:loss}
	\vspace{2mm}
	\includegraphics[width=0.625 \linewidth]{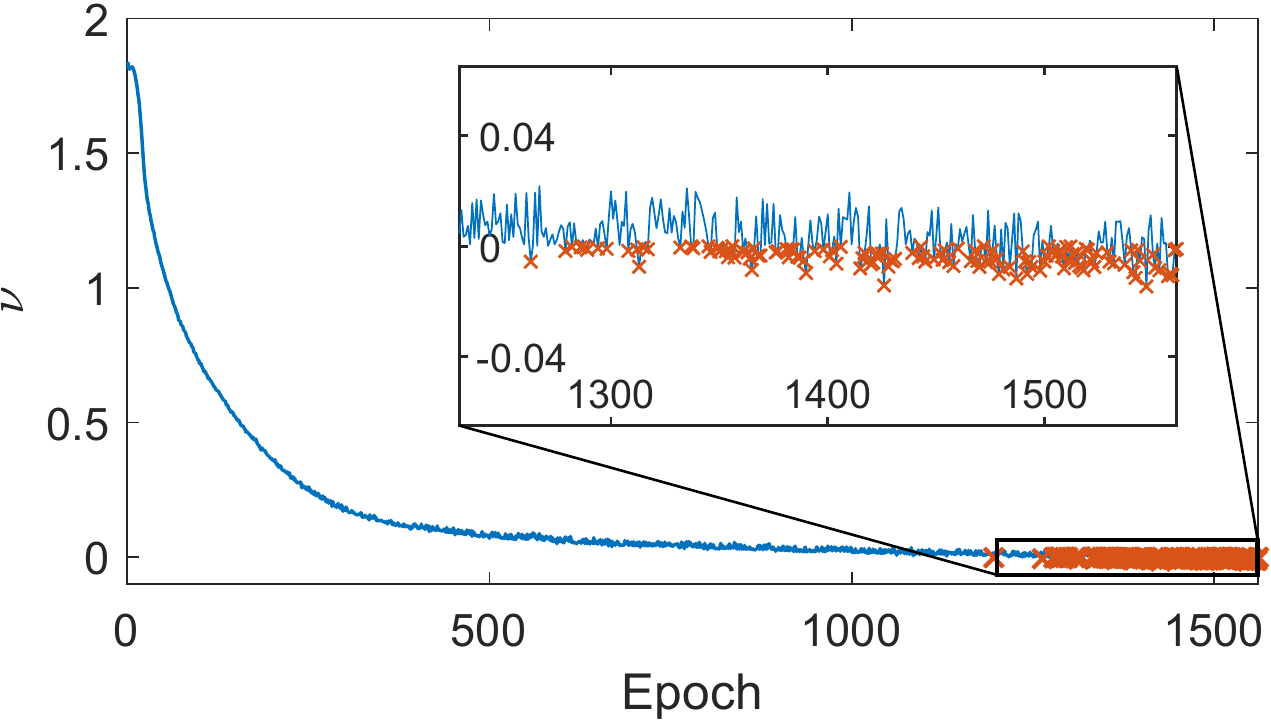}
	\vspace{-3mm}
	\caption{Evolution of the $\delta$ISS constraint residual $\nu$, throughout the training procedure. Red crosses indicate $\nu < 0$, i.e. that condition \eqref{eq:iss:condition} is satisfied.}
	\label{fig:residual}
\end{figure}

We adopted RMSProp as training algorithm to minimize the loss function $L$, using single trajectories as batches. 
The evolution of the loss function $L$ throughout the training procedure is depicted in Figure \ref{fig:loss}, whereas Figure \ref{fig:residual} shows the evolution of the stability constraint's residual $\nu$.
An early stopping rule was implemented to interrupt the training when the modeling performances on the validation set stop improving, so as to avoid overfitting.  
The training took $1560$ epochs, and led to $\| U_0 \| = 0.453$ and $\| U_1 \| = 0.985$. 
The corresponding stability constraint residual is $\nu \approx -0.001 < 0$, implying that the trained NNARX is both ISS and $\delta$ISS.

\begin{figure}
	\centering
	\includegraphics[width=0.75 \linewidth, clip, trim=0.0cm 0cm 0.07cm 0cm]{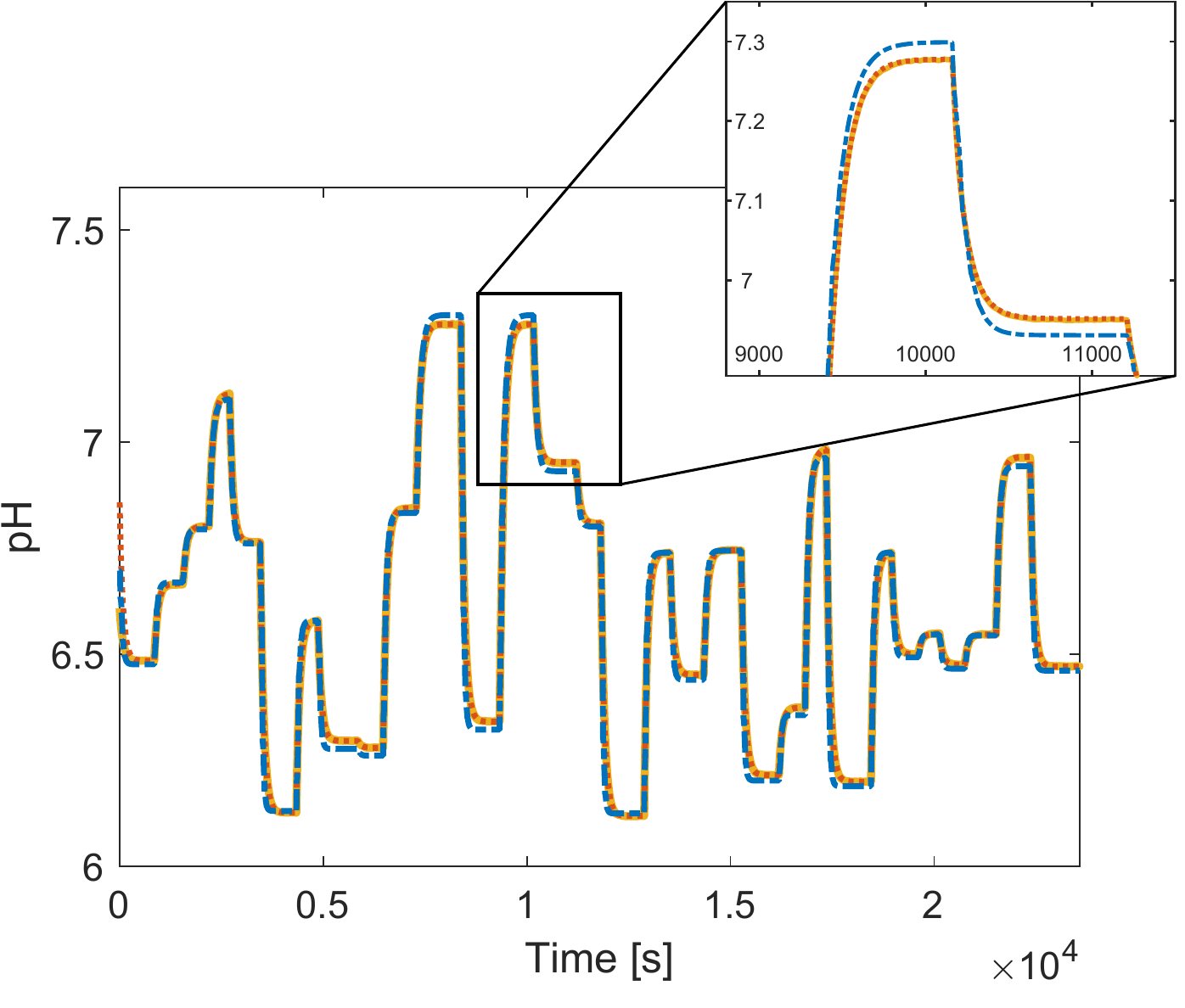}
	\vspace{-4mm}
	\caption{Performances of the trained model on the independent test set. 
		NNARX prediction (blue dashed line) compared to the ground truth (red dotted line) and to the LSTM prediction \citep{bonassi2019lstm} (yellow solid line). Note that the LSTM prediction and the ground truth are almost overlapping.}
	\label{fig:test_lstm}
\end{figure}

Eventually, the trained model was tested on the independent test-set used in \cite{bonassi2019lstm}, to validate the NNARX modeling performances in open-loop simulation, and to compare these performances with those of LSTMs.
The results are shown in Figure \ref{fig:test_lstm}.
To objectively evaluate the performances, we introduce the FIT index, defined as 
\begin{equation}
	\text{FIT} = 100 \cdot \left( 1 - \frac{\| \bm{y} - \bm{y}^{\{ts\}} \|}{\| \bm{y}^{\{ts\}} - \bar{y} \|} \right),
\end{equation}
where $(\bm{u}^{\{ts\}}, \bm{y}^{\{ts\}})$ denote the input and output sequence of the test set, $\bm{y}$ is the corresponding open-loop NNARX prediction, and $\bar{y}$ is the mean value of $\bm{y}^{\{ts\}}_k$.
The NNARX model scored $\text{FIT} = 91\%$, which is indeed satisfactory, although this architecture turned out to be slightly less accurate than LSTMs, which scored $\text{FIT} = 98 \%$.
This is definitely expected, given that NNARXs have a significantly simpler structure with respect to LSTMs, which comes at the price of lower representational capabilities. 
Better performances are expected when NNARXs are operated in closed-loop and re-initialized at each time instant using the past measured data.

\section{Conclusion}
In this paper we have studied the stability properties of Neural NARXs (NNARXs), i.e. discrete-time NARXs where the output regression function is a feed-forward neural network.
In particular, a sufficient condition for the Input-to-State Stability and for the Incremental Input-to-State Stability has been stated in the form of an inequality on network's weights.
NNARXs have then been tested on the \emph{pH} neutralization process benchmark system, showing satisfactory modeling performances.
Moreover, a preliminary comparison of NNARXs' performances with those of more complex neural networks shows that  their simple structure leads to a limited performance degradation.
A more extensive comparison with other neural networks' architectures will be subject of future research work.

\bibliography{NARX_stability_SYSID21}
\end{document}